%% file: issues-1col.tex
\documentclass[11pt,letterpaper]{article}
\usepackage{times}
\usepackage{mysetup,xspace}

\FULLPAGE

\newcommand{\fakeItem}[1][$\bullet$]{\vspace{2mm}{\bf #1}~~}

\usepackage[numbers]{natbib}
\newcommand{\citenoun}[1]{\citet{#1}}

\newcommand{\citeKarger}{\citet{KOS11,KOS13}\xspace}

\newcommand{\citeDynProc}{\citet{DynProcurement-ec12,BwK-focs13,BwK-SCUGC-13}\xspace}

\newcommand{\citeBwK}{\citet{BwK-focs13,BwK-SCUGC-13}\xspace}

\title{{\sc position paper}\vspace{2mm}\\
Online Decision Making in Crowdsourcing Markets:\\
Theoretical Challenges}

\author{Aleksandrs Slivkins%
\thanks{Microsoft Research. Email: {\tt slivkins@microsoft.com}.}
\and
Jennifer Wortman Vaughan%
\thanks{Microsoft Research. Email: {\tt jenn@microsoft.com}.}
}

\date{May 2013 \\ This version: November 2013}
\begin{document}
\maketitle

\begin{abstract}
\input{sec-abstract.tex}
\end{abstract}

\input{sec-intro.tex}
\input{sec-choices.tex}

\input{sec-specific.tex}

\input{sec-conclusions.tex}

\input{sec-acks.tex}

% ############ BIBLIOGRAPHY ##############

\begin{small}
\bibliography{bib-abbrv,bib-slivkins,bib-AGT,bib-bandits,bib-crowdsourcing-Alex,bib-taskassign,bib-PA}
\bibliographystyle{plainnat}
\end{small}

\end{document}

%% file: sec-abstract.tex
%\begin{abstract}
Over the past decade, crowdsourcing has emerged as a cheap and efficient method of obtaining solutions to simple tasks that are difficult for computers to solve but possible for humans.  The popularity and promise of crowdsourcing markets has led to both empirical and theoretical research on the design of algorithms to optimize various aspects of these markets, such as the pricing and assignment of tasks.  Much of the existing theoretical work on crowdsourcing markets has focused on problems that fall into the broad category of online decision making; task requesters or the crowdsourcing platform itself make repeated decisions about prices to set, workers to filter out, problems to assign to specific workers, or other things.  Often these decisions are complex, requiring algorithms that learn about the distribution of available tasks or workers over time and take into account the strategic (or sometimes irrational) behavior of workers.

As human computation grows into its own field, the time is ripe to address these challenges in a principled way.  However, it appears very difficult to capture all pertinent aspects of crowdsourcing markets in a single coherent model.  In this paper, we reflect on the modeling issues that inhibit theoretical research on online decision making for crowdsourcing, and identify some steps forward.  This paper grew out of the authors' own frustration with these issues, and we hope it will encourage the community to attempt to understand, debate, and ultimately address them.

%\bf{The authors welcome feedback for future revisions of this paper.}
%\end{abstract}

%% file: sec-intro.tex
\section{Introduction}

Crowdsourcing markets have emerged as a tool for bringing together \emph{requesters}, who have tasks they need accomplished, and \emph{workers}, who are willing to perform these tasks in a timely manner in exchange for payment.  Some crowdsourcing platforms, such as Amazon Mechanical Turk, focus on small ``microtasks'' such as labeling an image or filling out a survey, with payments on the order of ten cents, while other platforms, such as oDesk, focus on larger jobs like designing websites, for significantly larger payments, but most share the common feature of \emph{repeated interaction}. In these markets, workers, requesters, and the platform itself can all adjust their behavior over time to adapt to the environment.  For example, requesters can choose which workers to target (or filter out) for their tasks and which prices to offer, and can update these choices as they learn more about salient features of the environment, such as workers' skill levels, the difficulty of their tasks, and workers' willingness to accept their tasks at given prices.  In principle, the platform could also make smarter decisions over time as it learns more about the quality of both requesters and workers.

Naturally, as crowdsourcing has gained popularity and human computation has grown into its own field, researchers have taken an interest in modeling and analyzing the problem of online decision making in crowdsourcing markets.
There is plenty of prior work on which to build.  Online decision algorithms have a rich literature in operations research, economics, and several areas of computer science including machine learning, theory of algorithms, artificial intelligence, and algorithmic mechanism design. A large portion of this literature is concerned with the tradeoff between \emph{exploration} (obtaining new information, perhaps by sacrificing near-term gains) and \emph{exploitation} (making optimal decisions based on the currently available information). This tradeoff is often studied under the name \emph{multi-armed bandits}; a reader can refer to~\citenoun{CesaBL-book}, \citenoun{Bergemann-survey06}, \citenoun{Gittins-book11}, and \citenoun{Bubeck-survey12} for background and various perspectives on the problem space.
Another (somewhat overlapping) stream of work concerns \emph{dynamic pricing} and, more generally, \emph{revenue management} problems; see~\citenoun{BZ09} for an overview of this work in operations research, and~\citenoun{DynPricing-ec12} for a theory-of-algorithms perspective. Further, there is extensive literature on online decision problems in which all information pertinent to a given round is revealed, either before or after the algorithm makes its decision for this round;
see \citenoun{Borodin-book98}, \citenoun{Buchbinder-survey09}, and~\citenoun{CesaBL-book} for background.

Despite the vast scope of this existing work, crowdsourcing brings an array of domain-specific challenges that require novel solutions.  To address these challenges in a principled way, one would like to formulate a unified collection of well-defined algorithmic questions with well-specified objectives, allowing researchers to propose novel solutions and techniques that can be easily compared, leading to a deeper understanding of the underlying issues.  However, it appears very difficult to capture all of the pertinent aspects of crowdsourcing in a coherent model.  As a result, many of the existing theoretical papers on crowdsourcing propose their own new models.  This makes it difficult to compare techniques across papers, and leads to uncertainty about which parameters or features matter most when designing new platforms or algorithms.

There are several reasons why developing a unified model of crowdsourcing is difficult.  First, there is a tension between the desire to study models based on existing platforms, such as Mechanical Turk, which would allow algorithms to be implemented and tested immediately, and the desire to look ahead and model alternative platforms with novel features that could potentially lead to improved crowdsourcing markets further down the line.  A plethora of different platform designs are possible, and different designs lead to both different models and different algorithmic questions.

Second, even after a model of the platform has been determined, one must take into account the diversity of tasks and workers that use the platform.  Some workers may be better at some tasks than others, and may find some tasks harder than others.  It is natural to take this diversity into account when designing algorithms (especially for a problem like task assignment), but the way in which this diversity is modeled may impact the choice of algorithm and the theoretical guarantees that are attainable.

Third, crowdsourcing workers are human beings, not algorithmic agents (with the possible exception of spambots, which bring their own problems).  They may act strategically, maximizing their own welfare in response to the incentives provided by the requesters and the platform. Taking into account this strategic behavior is essential. Moreover, workers also may act in ways that are even less  predictable and seemingly irrational.  As just one example, empirical studies have shown that workers may not know their own costs of completing a task, and are susceptible to the \emph{anchoring effect} in which they judge the value of a task based on the first price they see (see~\citenoun{MW09}, \citenoun{MS12}, \citenoun{YCS13} and the literature reviews therein). One must be aware of these nuances when modeling worker behavior.

This paper offers a detailed reflection on the modeling issues that inhibit theoretical research on repeated decision making in crowdsourcing.  Our goal is not to offer a single unified model, but to raise awareness of the issues and spark a discussion in the community about the best ways to move forward.  To this end, we identify a multitude of modeling choices that exist, and describe a few specific models that appear promising in that they potentially capture salient aspects of the problem.

%These lists are not meant to be exhaustive, and we hope that as the community of researchers interested in the
%theory of crowdsourcing markets grows, they will be expanded and some consensus will be reached on the
%problems and models that are of the most importance.

\xhdr{A note on our scope.}
There have been some recent empirical or applied research projects aimed at developing online decision making algorithms that work well in practice on existing crowdsourcing platforms, primarily for the problems of task assignment and label aggregation (see, for example, \citet{CLZ13}, \citet{I+13}, and \citet{I+13b}).  While it is of course valuable for anyone interested in theoretical models of crowdsourcing to be aware of related empirical advances, we do not attempt to provide a complete survey of this work here. 

A similar attempt has been made to categorize future directions of crowdsourcing research within the human-computer interaction and computer-supportive cooperative work communities~\cite{K+13}.  While their paper discusses many of the same broad problems as ours (such as quality control, platform design, and reputation), the research agenda they lay out is heavily focused on the high-level design of the technical and organizational mechanisms surrounding future crowd work systems, while we focus more specifically on low-level modeling and algorithmic challenges for online decision making.

\section{Informal problem statement}

Let us start with an informal and very general description of the class of problems that fall under the umbrella of online decision making in crowdsourcing markets. There are three parties: workers, requesters, and the crowdsourcing platform. Over time, requesters submit tasks to the platform. Workers get matched with requesters, perform tasks, and receive payments. Workers and requesters may arrive to the platform and leave over time. Some workers may be better than others at some tasks; some tasks may be more difficult than others for some workers.  Some workers may enjoy some tasks more than others. All parties make {\bf repeated decisions} over time. All parties can {\bf learn over time}, which may help them in their decision making.  All decision makers receive {\bf partial feedback}: they see the consequences of their decisions, but typically they do not know what would have happened if they had made different decisions. Workers and requesters can {\bf behave strategically}, so their incentives need to be taken into account. The problem is to design algorithms for decision making, on behalf of the platform, the requesters, or perhaps even the workers.

We can think about this setting from the point of view of each of the three parties, who face their own choices and have differing motivations and goals:

\begin{itemize}

\item Requesters can choose the maximal price they are willing to offer for a given task, and specify any budget constraints that they have. They may also be able to choose which prices to offer to given (categories of) workers for given (categories of) tasks. Further, they may be able to choose which (categories of) workers they are willing to interact with, and which instances of tasks to assign to each (category of) worker.  The utility of the requester is typically (an increasing function of) the value that he obtains from completed work minus the price that he has to pay for this work.

\item The platform may match (subsets of) requesters to (subsets of) workers. In principle, the platform may also be able to modify the offered prices, within constraints specified by the requesters, and to determine how to charge requesters and/or workers for their use of the platform.  In the long run, the platform cares about maximizing long-term revenue.  In the short term, the platform may care about keeping workers and requesters happy to attract more business, especially in the presence of competing platforms.

\item Workers can decide whether to accept a task at a given price, and how much effort to put into it. They may be able to choose between (categories of) tasks, and may be asked to provide some information such as their interests, qualifications, and asking prices.  Workers care about the amount of money that they earn, the cost of their labor (in terms of time or effort), and perhaps the amount of enjoyment they receive from completing tasks.
\end{itemize}

As parties interact over time, they may be able to learn or estimate workers' skill levels, how much workers value their effort and time, the difficulty of tasks, and the quality level of particular workers on particular tasks.

We focus on two natural versions of the problem: \emph{requester-side} and \emph{platform-side}.\footnote{In principle, one could also consider the decision making problems faced by workers.  This area is still largely unexplored and could be a source of interesting research questions.  However, we do not consider it here.}  The requester-side problem is to design a mechanism which makes repeated decisions for a specific requester. The platform-side problem is to design a mechanism which makes repeated decisions for the platform. Both versions should take into account realistic constraints and incentives of workers and requesters.

%% file: sec-choices.tex
\section{Possible modeling choices}
\label{sec:choices}

To study and understand any particular aspect of repeated decision making for crowdsourcing markets, one must first determine a model or framework in which to work.  To date, most theoretical papers on crowdsourcing have proposed their own models, which are often hard to compare, and with good reason; one could come up with any number of valid models, and each has its own appeal.  However, we would argue that in order to extract and solve the fundamental issues that are present in any crowdsourcing market, one must be keenly aware of the (implicit or explicit) modeling choices they make, and the extent to which their algorithmic ideas and provable guarantees are sensitive to their modeling assumptions --- essentially how \emph{robust} results are to modifications in the model.

In this section, we identify a variety of modeling choices for repeated decision making in crowdsourcing, some of which have been discussed in prior work and some of which have not.  We break these choices roughly into five categories: task design, platform design, quality of work, incentives and human factors, and performance objectives, though other categorizations are possible.

\xhdr{Task design.} A wide variety of tasks can be crowdsourced. In many tasks, such as translation~\cite{Zaidan-acl11}, audio transcription~\cite{scribe-w4a13}, image/video analysis \cite{vizwiz,platemate-uist11}, or trip planning~\cite{mobi-hci12}, the workers' output is unstructured. On the other hand, in many applications currently deployed in the Internet industry, the workers' output has a very specific, simple structure such as a multiple-choice question or a free-form labeling question. In particular, most tasks that the assess relevance of web search results are structured as multiple-choice questions. The output structure (or lack thereof) determines whether and how workers' output can be aggregated. Below we survey some possibilities in more detail.

\fakeItem {\em Multiple-choice questions.} The simplest structure is a binary question (\emph{Is this a correct answer? Is this spam?}). Alternatively, a question can include more than two options (\emph{Which of these web search results is more appropriate for a given query? Which of these categories is more appropriate for this website?}) or allow workers to submit more than one answer when appropriate. Further, the possible answers may be numeric or otherwise ordered (\emph{What age group does this person belong to?}), in which case the similarity between the answers should be taken into account.

\fakeItem  {\em Free-form labeling.} In image labeling and various classification tasks, a worker may be allowed to submit arbitrary labels or categories. A worker may be allowed to submit more than one label for the same task. Some task designs may include a list of prohibited labels so as to obtain more diverse results.

\fakeItem  {\em Rankings and weights.} Some tasks strive to compute more complex structures such as rankings or relative weights for a given pool of alternatives~\cite{Pfeiffer-aaai12,Mao-aaai13}.  Each worker may be asked to provide either a full ranking or weight vector or only a comparison between a given pair of alternatives.

\xhdr{Platform design.}  All interaction between workers and requesters takes place in the context of a particular crowdsourcing platform. The platform designer controls the way in which requesters and workers may interact.  In modeling crowdsourcing markets, one may wish to model the features of a specific existing platform (such as Amazon Mechanical Turk, oDesk, or TaskRabbit) or explore alternative platform designs that may lead to improved crowdsourcing markets in the future.  In either case, one must consider the following modeling issues.

%\begin{itemize}

\fakeItem {\em Selection of workers.} Does the platform allow requesters to limit their tasks to specific (categories of) workers, and if so, how?
%
%Most current platforms do not allow little or no requester-side selection, but this can change in the near future.
%
Possibilities range from coarse platform-defined categories of workers (e.g., filtering by demographic information or feedback rating/reputation) to allowing requesters to specify arbitrary workers by worker ID.

\fakeItem {\em Selection of tasks.} How does the platform allow workers to select tasks? Does a worker see \emph{all} offerings that are currently in the market?  If so, how are they sorted?  Alternatively, the platform could limit the options presented to a given worker. When the task(s) from a given requester are offered to a given worker, does the platform allow the requester to restrict which offerings from other requesters this worker is exposed to?

\fakeItem {\em Feedback to the requester.} What is the granularity of feedback provided to a given requester by the platform? In particular, does the platform expose the workers' performance by individual IDs or only coarse categories? Does the platform expose some or all of the workers' attributes? Does the platform expose workers' performance for other requesters?  If a worker views a task but ultimately does not decide to complete it, does the requester of that task learn that it was viewed?

\fakeItem {\em Temporal granularity.} How often does the platform allow a requester to adjust its selection preferences? How frequently does the platform provide feedback to requesters?  When is it reasonable to assume that workers complete a given task sequentially, one at a time, and when is it necessary for a model to allow multiple tasks to be accepted or completed simultaneously?

\fakeItem {\em Price negotiation.} How do requesters negotiate prices with workers?  Typically it is assumed that requesters post one take-it-or-leave-it price, but this is not the only possibility.  Can a requester post different prices to different workers depending on the workers' attributes? Can the requester update the price over time? Alternatively, one could imagine a crowdsourcing platform on which workers communicate minimal asking prices to the requesters and prices are determined via an auction.

\fakeItem {\em Payments.} How do payments happen? The simplest version is that each worker, upon completion of a task, receives the posted price for that task. Can a requester specify a payment that depends on the quality of submitted work? If so, must he declare the exact payment rule up front? In particular, must he commit to a specific way of determining the quality of submitted work?

%\xhdr{Platform-wide tracking of workers.}

\fakeItem {\em Persistent identities.} Does a platform ensure persistent worker identities? How difficult or expensive is it for a worker to abandon his old identity and create a new one, or to create multiple identities?

\fakeItem {\em Platform-wide reputation.} Does the platform maintain a platform-wide reputation score for each worker and/or requester? If so, then how exactly is this score defined?  What properties of this score are guaranteed?  How are reputations used?  How are workers and requesters prevented from manipulating their own reputations and the reputations of others?

\fakeItem {\em Qualifying tests.}  Does the platform allow requesters to require workers to pass a qualification test before being allowed to work for money?

%\end{itemize}

\xhdr{The quality of work.}  In crowdsourcing environments, both tasks and workers are highly diverse.  Naturally certain workers will be able to complete certain tasks more proficiently than others, yielding higher quality results, possibly for less effort.  Although the difficulty of a task and the skills of each worker are generally not known to the platform or requester a priori, they can be learned and refined over time, though this is made more difficult when the quality of a completed task is not observable or easily measured, or when the users of the market are themselves changing over time.

%\begin{itemize}

\fakeItem {\em Worker skills and task difficulty levels.}  Some workers may be better than others at some tasks, and worse than others at other tasks.  Ideally, an online algorithm should learn estimates of workers' skills for various tasks over time.  However, if each worker has an arbitrary skill level for each task, it will be impossible to generalize based on observed performance.  Therefore, it is desirable to model some type of structure on tasks or workers. Can the inherent skill level of a given worker be summarized by a single number? Can the inherent difficulty of a task be summarized by a single number? More generally, how should one model a given worker's proficiency at a given task?

\fakeItem {\em Effort.} The quality of a worker's output may also depend on the amount of effort that the worker puts into the task.   How can one model the effect of varying levels of effort on the quality of completed work?

\fakeItem {\em Time dependence.} Workers' skills and the quality of their completed work may change over time, either because of learning (perhaps after completing the same task several times) or changes in effort level.  Additionally, properties of the population of workers (such as the size of the population or their average quality) may change over time, for example based on time of day.

\fakeItem {\em Measurable vs. unmeasurable quality.}  Some tasks, such as labeling an image as containing a face or not, or answering a multiple-choice question, have objective measures of quality; either the provided answer is correct, or it is not.  For other tasks, such as translating a paragraph from one language to another or drawing a picture, the quality of the work is more difficult to quantify.  For tasks without objective measures of quality, it is necessary to clearly define criteria on which a task will be judged.

\fakeItem {\em Observable vs. unobservable quality.}  Even if the response to a task is objectively right or wrong, as is the case in the image labeling task above, the quality of submitted work may not be immediately observable; if the requester already knew the correct label, he wouldn't need the worker to provide it.  For many tasks of this form, it is possible for the requester to assign the same question to multiple workers and estimate the correct answer by consensus; if a statistically significant (weighted) majority agrees on some answer, it may be safe to assume it is correct.

\fakeItem {\em Availability of gold standard tasks.}  In some cases in which task quality is not observable, a requester may have access to a set of ``gold standard" tasks for which the correct answer is known a priori.  He can then use these tasks to bootstrap the process of learning workers' skill levels (possibly as part of a separate qualifying test, if allowed).

%\end{itemize}

\OMIT{
\begin{OneLiners}
\fakeItem the quality of answers is immediately observable for all tasks.

\fakeItem we have a limited supply of ``gold HITs".

\fakeItem the correct answers are not known a priori, but can be estimated via consensus.

\end{OneLiners}
}

\xhdr{Incentives and other human factors.}  Perhaps the most difficult and controversial aspect of formalizing the repeated decision making problem in a crowdsourcing environment is modeling the incentives of participants.  In the economics literature, it is standard to model agents as self-interested and rational utility maximizers.  Empirical research studying the behavior of workers on Mechanical Turk suggests that there may be difficulties applying such models in crowdsourcing settings in which workers do not have a good sense of their own costs of completing a task and may choose particular tasks based on intangible qualities such as fun.  This leads to another set of modeling issues.

%\begin{itemize}

\fakeItem {\em Rationality.}  Do workers behave rationally when choosing tasks to complete at particular prices? The standard model from economics is that each worker maximizes a simple well-defined notion of ``utility," (an increasing function of) payment received minus production cost. However, workers may also take into account other factors, such as perceived value to society and personal enjoyment. The significance of these intangible factors in crowdsourcing environments is not well-understood.

\OMIT{Can the workers strategically alter their performance after committing to a particular task, for example by choosing the amount of effort that they will put into the task?}

\fakeItem {\em Effort levels.} Can the workers strategically alter the amount of effort that they put into a given task, or is it assumed that workers complete each task to the best of their ability?

\fakeItem {\em Costs.}  How do the workers evaluate their internal costs for completing a given task or putting forth a given level of effort? Is this internal cost even a well-defined quantity? It may, in principle, depend on subjective perceptions (such as value to society and personal enjoyment), and also on the offered prices (the so-called \emph{anchoring effect}~\cite{TK74,MW09}). Even if the internal cost is a well-defined quantity, do workers know it?

\OMIT{ %%%%
\item {\em Other biases.} Even if internal costs are well-defined and known to each worker, how do the workers decide whether or not to accept a particular task at a given price?  Do they simply maximize utility?  Workers may have a perception of a ``fair price" which is separate from their internal costs, and may be reluctant to accept a task at a price below the perceived ``fair price.''
} %%%%%

\fakeItem {\em Fair price.} Workers may have a perception of a ``fair price" which is separate from their internal costs, and may be reluctant to accept a task at a lower price~\cite{MW09,YCS13}.

\fakeItem {\em Task-specific motivations.} In some cases, workers may be motivated by task-specific factors.  For example, if a task involves creating content to be posted on the web, workers may be motivated by the possibility of receiving attention~\cite{GH13}.

\fakeItem {\em Myopia vs. long term strategies.}
How myopic are the workers? When making a decision about a particular task, do they take into account the effects of this decision on their future? For example, a worker may be reluctant to accept highly priced tasks that she is not suitable for, because her bad performance on these tasks would impact her reputation score with the platform and/or the requester, and therefore affect what she is offered in the future. For a similar reason, a worker may accept an easy but low paying task, or exert additional effort. More generally, workers may be able to strategically alter their behavior in order to affect the long-term behavior of a particular requester or the platform.

\fakeItem {\em Strategic timing.} Can the workers strategically choose the time intervals during which they are online? If a requester is using an online learning algorithm, some phases of this algorithm may be more beneficial for workers than others. For example, if a requester starts out with a low price for a given category of tasks, and then gradually adjusts it upwards, then workers may want to arrive later rather than sooner.

\xhdr{Performance objectives.}
The statement of an algorithmic problem should include a specific performance objective to be optimized, perhaps under some constraints.

A requester-side decision making algorithm typically maximizes the utility of this requester, i.e., the value of completed tasks minus the payment to the workers.  If the requester has a pre-specified budget, the algorithm may instead maximize the value of completed tasks subject to budget constraints.  In some settings it may be feasible to take into account the workers' happiness, so as to encourage them to stay with this requester in the future. Specifically, in incentive-compatible mechanisms that elicit workers' production costs one may wish to optimize the \emph{social welfare} -- the utility of the requester plus the total utility of the workers -- or some other weighted sum of these two quantities.

For platform-side algorithms, the performance objective might take into account the utility of requesters, workers, and the platform itself. Assuming the platform receives a fixed percentage of every transaction, the platform's revenue is simply a fixed fraction of the total payment from requesters to workers. One reasonable objective is a weighted sum of the platform's revenue and the total utility of the requesters; the choice of weights is up to the platform. Moreover, the platform may wish to ensure some form of fairness, so that no requesters are starved out. Further, the platform may have additional forward-looking objectives that are not immediately comparable with the platform revenue and requesters' utility, such as the market share and the workers' happiness (to encourage workers to stay with the platform in the future).

%% file: sec-specific.tex
\section{Specific directions}
\label{sec:models}

Given the abundance of modeling choices identified, it appears hopeless to seek out a single unified model that attempts to capture all aspects of repeated decision making in crowdsourcing markets.  Nevertheless, it may be possible to study, understand, and solve the fundamental issues that arise in this problem space by carefully choosing specific (families of) models that encapsulate specific salient features  --- especially if we, as a community, keep generalizability and robustness in mind as we make our modeling and algorithmic choices.

Below we identify and discuss several directions in the problem space that appear ripe for near-term progress. As it happens, these directions encompass most existing work of which we are aware.

\subsection{Adaptive task assignment}

One problem of interest is assigning tasks to workers with the goal of maximizing the quality of completed tasks at a low price or subject to budget constraints. In this \emph{task assignment} problem, strategic issues are ignored in order to gain analytical tractability; the model typically does not touch on the way in which prices are set, and does not include workers' strategic responses to these prices.  It can be studied in settings in which the quality of work performed is immediately observable
%(as might be the case for retrieval tasks, such as locating images of people engaged in a particular image)
or settings in which it is not.

Much of the existing work on task assignment focuses on \emph{classification tasks}, in which workers are asked multiple-choice questions to provide labels for images, websites, web search results, or other types of queries. Then it is natural to assume that the quality of performed work (i.e., the correct label) is not immediately observable; instead, the requester may have a limited supply of gold standard tasks that can be used to learn more about workers' skills.  While the problem of inferring the solutions to classification problems using labels from multiple sources has been studied for decades in various forms~\cite{DS79,SPI08,CKW05,CKW08,DS09,Welinder-nips10}, here we focus on the problem of choosing workers in an online fashion to provide the labels; this problem is relatively new.

Existing research on task assignment typically focuses on the problem faced by a single requester.  Each worker charges a fixed and known cost per task completed (which may or may not be the same for all workers), and each task may be assigned to multiple workers in order to the improve the quality of the final result (as is necessary for classification tasks). The objective is to optimize a tradeoff between the total cost and the quality of the completed tasks, for example, minimizing the total cost of work subject to the quality being above a given threshold. Partial information on the task difficulty and the workers' skills may or may not be known, e.g., in the form of reputation scores or Bayesian priors.

In the most common variant of this problem, workers arrive online and the requester must assign a task (or sequence of tasks) to each new worker as she arrives.
\citeKarger
introduced one such model for classification tasks and proposed a non-adaptive assignment algorithm based on random graph generation along with a message-passing inference algorithm inspired by belief propagation for inferring the correct solution to each task.  They proved that their technique is order-optimal in terms of budget when each worker finds all tasks equally difficult. Other models of this form have been studied both for tasks with observable quality~\cite{HV12} and for classification tasks with unobservable quality but access to gold standard tasks~\cite{HJV13}.  In these papers, the authors utilize online primal-dual techniques \cite{Buchbinder-survey09,Devanur11}
%~\cite{BN05,BN06,DevanurH09,Devanur11}
and show that adaptive task assignment yields an improvement over non-adaptive assignment when the pool of available workers and set of tasks are diverse.

Alternatively, one might consider a model in which the requester may choose a particular worker or category of workers to assign to a particular task rather than choosing a task to assign to a particular worker.  This setting already encompasses a variety of specific models of varying levels of difficulty, but even a version with a single task assigned multiple times is quite challenging \cite{BanditSurveys-colt13}.

It is worth noting that the models described above are not covered by prior work on multi-armed bandits (where an algorithm observes a reward after each round and the goal is to maximize the total reward over time). Essentially, this is because in adaptive task assignment each rounds brings \emph{information} rather than reward; the value of this information is not immediately known, and in any case the goal is not to maximize the total value of collected information but to arrive at high-quality solutions for the tasks.

On the other hand, multi-armed bandit formulations are more directly applicable to versions of adaptive task assignment in which the quality or utility  of each completed task is immediately observable, and the goal is to maximize the total utility over time. The basic version in which the tasks are homogeneous and the algorithm chooses among workers has been studied in \citet{TranThanh-ecai12}; the novelty here is that one needs to incorporate budget constraints. This version falls under the general framework that was later defined and optimally solved in \citet{BwK-focs13}.

The problem formulations surveyed above are quite different from one another, and the algorithms and theoretical guarantees are therefore incomparable.  There is a lot of room for further work extracting the key features of the task assignment problem and take-away messages that generalize to a wider range of models and assumptions.

\OMIT{ %%%%%%%
The models in these papers differ, and the algorithmic results and theoretical guarantees are therefore incomparable.  While one would like a unified model of task assignment that allows for easier comparisons, some useful generalizations have come out of these papers.  For example, prior theoretical work \cite{KOS11a,KOS11} suggested that in some circumstances, the assignment algorithm does not matter in the sense that optimizing assignments does not yield an advantage over random assignment.  The papers mentioned above illustrate that when the crowd of workers is diverse, the assignment algorithm can make a difference, and therefore should be optimized when possible.
Still, there is a lot of room for further work extracting the key features of the task assignment problem and take-away messages that generalize to a wider range of models and assumptions.
}

\OMIT{
One can also study a more involved, platform-side version, where in each round the platform needs to route tasks from multiple requesters. The goal would be to minimize the total cost, over all requesters, under constraints on the quality of  completed tasks and fairness (non-starvation) over requesters.

Several recent papers targeted this problem space [cite], but, to the best of our understanding, there is a lot of room for further work, both in terms of generality of settings and algorithm performance in settings that have already been studied.  Still, there is a lot of room for further work, both in terms of generality of settings and algorithm performance in settings that have already been studied.
}

\subsection{Dynamic procurement}

Dynamic procurement focuses on repeated \emph{posted pricing}, as applied to hiring workers in a crowdsourcing market.%
\footnote{The term ``dynamic procurement" is from
\citeDynProc.
%Badanidiyuru et al.~\citeyear{DynProcurement-ec12,BwK-focs13,BwK-SCUGC-13}.
}
Each interaction between the requester and the workers follows a very simple protocol: the requester posts a price (for a given task) which can be either accepted or rejected by the workers.

The basic model of dynamic procurement is as follows. In each round, the requester posts a task and a price for this task, and waits until either some worker accepts the task or the offer times out.  The objective is to maximize the number of completed tasks under fixed constraints on budget and waiting time.
A standard model of workers' rationality is assumed.  In particular, each worker has a fixed ``internal cost" which is known to the worker but not to the mechanism. A worker accepts a task only if the posted price for this task matches or exceeds her internal cost. If multiple such tasks are offered at the same time, the worker picks the one which maximizes her utility (offered price minus the internal cost). The requester typically has a very limited knowledge of the \emph{supply curve}, the distribution of workers' internal costs. However, the supply curve may be learned over time.

The basic model described above assumes several significant simplifications as compared to the discussion in Section~\ref{sec:choices}. First, workers cannot strategically manipulate their effort level in response to the offered prices. Second, workers are myopic, in the sense that they do not take into account the effects of their decision in a given round on the future rounds. A standard way to model myopic workers is to assume that the requester interacts with a new worker in each round. Third, the supply curve does not change over time. Finally, the issues of (explicit) task assignment are completely ignored.

Despite all the simplifications, this basic model is very non-trivial. Considerable progress has been made in a recent line of work \cite{DynProcurement-ec12,BwK-focs13,BwK-SCUGC-13,Krause-www13}, but it is not clear whether the current results are optimal. Further, multiple extensions are possible \emph{within} the basic model. For instance, the requester may be interested in multiple categories of tasks, and may offer more complicated ``menus" that allow workers to perform multiple tasks of the same type.%
\footnote{The general result in %\citet{BwK-focs13,BwK-SCUGC-13}
\citeBwK
%Badanidiyuru et al.~\citeyear{BwK-focs13,BwK-SCUGC-13}
applies to some of these extensions, but the particular guarantees are likely to be suboptimal.}

\OMIT{
A recent line of work \cite{DynProcurement-ec12,BwK-focs13,BwK-SCUGC-13,Krause-www13} has studied dynamic procurement under the basic objective (maximizing the number of tasks), in a highly idealized model in which in each round the requester interacts with a new worker. Even this setting is very non-trivial, and the current results for this setting are not necessarily optimal. Multiple extensions are possible, e.g., to multiple categories of tasks and to more complicated ``menus" that allow workers to perform multiple tasks of the same type.
\footnote{The general result in \cite{BwK-focs13,BwK-SCUGC-13}
%\citeauthor{BwK-focs13} (\citeyear{BwK-focs13,BwK-SCUGC-13})
applies to some of these extensions, but the particular guarantees are likely to be sub-optimal.}
}

One can also consider a platform-side version of dynamic procurement, in which the platform is in charge of setting posted prices for all requesters, under budget constraints submitted by the requesters or imposed by the platform. The most basic goal would be to maximize the total number of completed tasks over all requesters, perhaps under some fairness constraints.

It is worth noting that dynamic procurement is closely related to \emph{dynamic pricing}: repeated posted pricing for selling items. In particular, dynamic procurement on a budget with unknown supply curve has a natural ``dual" problem about selling items: dynamic pricing with limited inventory and unknown demand curve. The latter problem has received some attention (e.g., \cite{BZ09,DynPricing-ec12,BesbesZeevi-or12,BwK-focs13}). Moreover, \citet{BwK-focs13} consider a general framework which subsumes both problems.

Dynamic procurement can be compared to a more general class of game-theoretic mechanisms in which workers are required to explicitly submit their preferences (see, e.g.,  \citet{SM13}). Dynamic procurement is appealing for several reasons (as discussed, for example, in \citet{DynPricing-ec12} and \citet{ChawlaHMS10}). First, a worker only needs to evaluate a given offer rather than exactly determine her internal costs and preferences; humans tend to find the former task to be much easier than the latter. Second, a worker reveals very little information about themselves: only whether she is willing to accept a particular offer. Third, posted pricing tends to rule out the possibility that workers may strategically alter their behavior in order to manipulate the requester.  However, it is possible that more general mechanisms may achieve better guarantees in some scenarios.

\xhdr{Behavioral effects.}
In order to ensure that results are applicable to real-world markets, apart from addressing dynamic procurement under the traditional assumptions on workers' rationality, it is desirable to incorporate more complicated behaviors, such as perceived fair prices and the anchoring effect. Then the prices offered by the requester may influence the properties of the worker population, namely change workers' perception of fair prices and/or production costs. The challenge here is to model such influence in a sufficiently quantitative way, and design dynamic pricing algorithms that take this influence into account. However, the existing empirical evidence \cite{MW09,YCS13} appears insufficient to suggest a particular model to design algorithms for; further, even surveying the potentially relevant models is not easy. To make progress,
there needs to be a convergence of empirical work on modeling that is geared towards algorithmic applications, and algorithm design work that is fully aware of the modeling issues.

One can draw upon a considerable amount of prior work on reference prices~\cite{Tversky-91,Putler-92,Kalyanaram-95,Thaler-08}, and some recent work on dynamic pricing with reference price effects~\cite{Popescu-OR07,Popescu-OR11}. However, all of this work is usually in the context of selling items to consumers, and therefore may not be directly applicable to crowdsourcing. Moreover,~\citet{Popescu-OR07} and \citet{Popescu-OR11} assume a known demand distribution.

In view of the above, one wonders how significant the behavioral effects are in real-life crowdsourcing markets, and whether it is safe to ignore them in the context of dynamic procurement. To the best of our understanding, this issue is currently unresolved; while significant behavioral effects have been observed in specifically designed, very short-term experiments, it is not clear whether the significance is preserved in longer-running systems.

\newcommand{\repeatedPA}{repeated principal-agent\xspace}
\newcommand{\RepeatedPA}{Repeated principal-agent\xspace}

\subsection{\RepeatedPA problem}

Consider an extension of dynamic procurement to a scenario in which the workers can strategically change their effort level depending on the price or contract that they are offered. The chosen effort level probabilistically affects the quality of completed work, determining a distribution over the possible quality levels. Each worker is characterized by a mapping from effort levels to costs and distributions over the quality levels; this mapping, called worker's \emph{type}, is known to the worker but not to the mechanism. Crucially, the choice of effort level is not directly observable by the requester, and cannot (in general) be predicted or inferred from the observed output. The single-round version of this setting is precisely the \emph{principal-agent problem}~\cite{LM02}, the central model in  contract theory, a branch of Economics.%
\footnote{In the language of contract theory, the requester is the  ``principal" and workers are ``agents."}

Posted pricing is not adequate to incentivize workers to produce high-quality work in this setting since workers could exert the minimal amount of effort without any loss of payment. Instead, requesters may want to use more complicated contracts in which the payment may depend on the quality of the completed work. Note that contracts \emph{cannot} directly depend on the worker's effort level, as the latter is not known to the requester. The mechanism may adjust the offered contract over time, as it learns more about the worker population.

To make the problem more tractable, it would probably help to assume that workers are myopic, and that the quality of completed tasks is immediately observable. As in dynamic procurement, the issues of  task assignment are completely ignored.

Even with these simplifications, the \emph{\repeatedPA} setting described above is significantly more challenging than dynamic procurement. Essentially, this because the principal-agent problem is a vast generalization of the simple accept/reject interaction in dynamic procurement. The space of possible contracts and the space of possible types are much richer than their counterparts in dynamic procurement, and the mapping from an action to requester's expected utility is more difficult to characterize and analyze. In particular, contracts are not limited to mappings from observed outcomes to payments. In principle, the requester can specify a \emph{menu} of several such mappings, and allow a worker to choose among them.
(It is well-known that using such ``menus" one can maximize the requester's expected utility among all possible interaction protocols between a single requester and a single worker~\cite{LM02}.)

A notable additional complication is that it may be advantageous to reduce the action space -- the class of contracts that an algorithm considers. While reducing the action space may decrease the quality of the best feasible contract, it may improve the speed of convergence to this contract, and make the platform more user-friendly for workers. For example, one may want to reduce the number of items in the ``menu" described above (perhaps all the way to single-item menus), reduce the granularity of the quality levels considered, or restrict attention to human-friendly ``monotone'' contracts, in which higher quality levels always result in higher payments to the workers.

%\xhdr{Prior work and current status.}
Perhaps surprisingly, the rich literature in contract theory sheds little light on this setting. Most work focuses on a single interaction with an agent whose type is either known or sampled from a known distribution, see~\citet{LM02} for background. Some papers~\cite{BFN06,MND12} have studied settings in which the principal interacts with multiple agents, but makes all its decisions at a single time point. In~\citet{Sannikov08}, \citet{William09}, and \citet{Sannikov12}, the principal makes repeated decisions over time, but the requester interacts with only a single agent who optimizes the long-term utility. The agent's type is sampled from a known prior.

\citet{CG06} consider the basic version of the \repeatedPA problem --- with single-item ``menus" and no budget constraints --- adapt several algorithms from prior work to this setting, and empirically compare their performance. In an ongoing project, \citet{RepeatedPA-13} design algorithms with (strong) provable guarantees for the same basic version, drawing on the connection with some prior work \cite{LipschitzMAB-stoc08,xbandits-nips08,contextualMAB-colt11} on multi-armed bandit problems. However, their guarantees make significant restrictions on the action space. We are not aware of any other directly relevant work (beyond the work on dynamic pricing, which does not capture the unobservable choice of worker's effort level).

Progress on \repeatedPA problems might suggest improvements in the design of crowdsourcing platforms. In particular, it may inform the platform's choice of the action space. Currently even the basic version of the \repeatedPA problem is not well-understood.

%\cite{Nisan-ec13,BFN06}

\subsection{Reputation systems}

Persistent reputation scores for workers may help limit spam and encourage high quality work. Likewise, persistent reputation scores for requesters may encourage requesters to be more considerate towards the workers. Thus, one may want to design a stand-alone ``reputation system" which defines and maintains such scores (perhaps with the associated confidence levels), as a tool which can be used in different applications and in conjunction with different higher-level algorithms. Reputation systems may be designed to address issues related to either \emph{moral hazard} (when workers' effort is desirable but not immediately observable) or \emph{adverse selection} (when similar-looking workers may be different from one another, and it is desirable to attract workers who are more skilled or competent), or a combination of the two.

There is already a rich literature on reputation systems for online commerce, peer-to-peer networks, and other domains; see, for example, \citenoun{FRS07}, \citenoun{R+00}, or the seminal paper of \citenoun{D05}.  However, the basic models for reputation systems have several limitations when applied to crowdsourcing markets. First, there may be domain-specific design goals that depend on a particular application or (even worse) on a particular algorithm which uses the reputation scores. Then it may be necessary to design the reputation system and the platform's or requesters' algorithms in parallel, as one inter-related problem. Second, reputation systems  are typically designed separately from task assignment, which artificially restricts the possibilities and can therefore lead to sup-optimal designs. Third, whenever the properties of the worker population are not fully known, there is an issue of exploration: essentially, it may be desirable to give the benefit of a doubt to some low-rated workers, so as to obtain more information about them.

\citet{ZV12} and \citet{H+12} examine the problem of designing a reputation system specific to crowdsourcing markets, building on the idea of \emph{social norms} first proposed by \citet{K92}.
Social norms consist of a set of prescribed rules that market participants are asked to follow, and a mechanism for updating reputations based on whether or not they do.  They pair reputation systems with \emph{differential service schemes} that allocate more resources or better treatment to those with higher reputation in such a way that market participants have the incentive to follow the prescribed rules of the market.
The social norms designed by \citet{ZV12} and \citet{H+12} address only the moral hazard problem, ignoring issues related to adverse selection by making assumptions about known types, and do so on greatly simplified models that capture some salient features of crowdsourcing systems (such as the difficulty of perfectly evaluating the quality of work and the fact that the population may be changing over time) but ignore many others (such as the vast diversity of workers and requesters in real systems and the fact that workers may be learning over time).

There is significant room for further research studying more realistic models, looking simultaneously at moral hazard and adverse selection, and better defining what an ``optimal'' reputation system means in general.

%%%%%%
\subsection{One common theme: The exploration-exploitation tradeoff}

Most models surveyed in this section exhibit a tradeoff between \emph{exploration} and \emph{exploitation}, i.e., between obtaining new information and making optimal per-round decisions based on the information available so far \cite{CesaBL-book,Bergemann-survey06,Bubeck-survey12}. This tradeoff is present in any setting with repeated decisions and partial, action-dependent feedback, i.e., whenever the feedback received by an algorithm in a given round depends on the algorithm's action in this round. The explore-exploit tradeoff occurs in many areas, including the design of medical trials, pay-per-click ad auctions, and adaptive network routing; numerous substantially different models have been studied in the literature, to address the peculiarities of the various application domains.
%(A reader can refer to \citenoun{Bubeck-survey12} for more background).

An issue which cuts across most settings with explore-exploit tradeoff, including those arising in crowdsourcing markets, is whether to use a fixed schedule for exploration, or to adapt this schedule as new observations come in. In the former case, one usually allocates each round to exploration or exploitation (either randomly or deterministically, but without looking at the data), chooses uniformly at random among the available alternatives in exploration rounds, and maximizes per-round performance in each exploitation round given the current estimates for the unknown quantities. In the latter case, which we call \emph{adaptive exploration}, exploration is geared towards more promising actions, without sacrificing too much in short-term performance. Often there is no explicit separation between exploration and exploitation, so that in each round the algorithm does a little of both. Adaptive exploration tends to be better at taking advantage of the ``niceness" of the problem instance (where the particular notion of ``niceness" depends on the model), whereas non-adaptive exploration often suffices to achieve the optimal worst-case performance.

For a concrete example, consider a simple, stylized model in which an algorithm chooses among two actions in each round, each action $x$ brings a 0-1 reward sampled independently from a fixed distribution with unknown expectation $\mu_x$, and the goal is to maximize the total reward over a given time horizon of $100$ rounds. One algorithm involving non-adaptive exploration would be to try each action 10 times, pick the one with the best empirical reward, and stick with this action from then on. (However, while in this example choosing the best action in each exploitation round is trivial, in some settings this choice can be the crux of the solution~\cite{HJV13}.) On the other hand, one standard approach for adaptive exploration is to maintain a numerical score (index or upper confidence bound) for each action, which reflects both the average empirical reward and the sampling uncertainty, and in each round to pick an action with the highest score.

A closely related issue, endemic to all work on dynamic procurement (and dynamic pricing), is \emph{discretization} of the action space (in this case, the price space). Essentially, it is complicated to consider \emph{all} possible prices or price vectors, and instead one often focuses on a small but representative subset of thereof. For example, one may focus on prices that are integer multiples of some a priori chosen $\eps\in (0,1)$. While optimizing among the remaining prices is very simple in the most basic setting of dynamic procurement without budget constraints~\cite{Bobby-focs03}, it can be very difficult in more advanced settings~\cite{BwK-focs13,BwK-SCUGC-13,Krause-www13}.  Prior work on related, but technically different explore-exploit problems \cite{LipschitzMAB-stoc08,xbandits-nips08,contextualMAB-colt11} suggests that it may be advantageous to choose the discretization \emph{adaptively}, depending on the previous observations, so that the algorithm zooms in faster on the more promising regions of the action space.

Understanding these issues is crucial to the design of algorithms for repeated decision making in crowdsourcing systems.  At the same time, models designed specifically for the crowdsourcing domain often present new algorithmic challenges in explore-exploit learning (as is the case for most work on adaptive task assignment and dynamic procurement).  Techniques designed to address these challenges may be applicable more broadly  in other scenarios in which the exploration-exploitation tradeoff arises. (For example, dynamic procurement for crowdsourcing markets is one of the major motivating examples behind the general framework in \citet{BwK-focs13}.)

%% file: sec-conclusions.tex
\section{Conclusions}

Crowdsourcing is a new application domain for online decision making algorithms, opening up a rich and exciting problem space in which the relevant problem formulations vary significantly along multiple modeling ``dimensions."  This richness presents several challenges.  Most notably, it is difficult for the community to converge on any particular collection of models, and as a result, it is difficult to compare results and techniques across papers. Additionally, any particular modeling choices have inherent limitations that are sometimes hard to see. It is not clear a priori which limitations will impact the practical performance of algorithms, making claims about robustness difficult to achieve.

In this paper, we attempt to raise awareness of these issues. Towards this goal, we identify a multitude of possible modeling choices, and discuss several specific directions in which progress can be made in the near future.  We hope that this will spark discussion and aide the community as it moves forward.

%% file: sec-acks.tex
\section*{Acknowledgements}

We are grateful to Ashwinkumar Badanidiyuru, Yiling Chen, Chien-Ju Ho, Robert Kleinberg, and Ariel Procaccia for enlightening conversations, pointers to relevant literature, and useful feedback on earlier drafts of this paper.  We thank Yiling Chen, Winter Mason, and Duncan Watts for sharing their thoughts on the anchoring effect and other behavioral considerations that are often ignored.  Additionally, Slivkins wishes to thank Ittai Abraham, Omar Alonso, Vasilis Kandylas, Rajesh Patel, Steven Shelford, and Hai Wu for many conversations regarding the practical aspects of crowdsourcing markets.